\documentclass[amsmath, amssymb, aps, reprint, superscriptaddress, nofootinbib]{revtex4-2}

\usepackage{graphicx}
\usepackage{dcolumn}

\usepackage{bm}
\usepackage{physics}
\usepackage[svgnames]{xcolor}
\usepackage[colorlinks=true, linkcolor=blue, citecolor=blue, urlcolor=blue]{hyperref}
\usepackage{amsthm}
\usepackage{enumitem}

\usepackage{docmute}

\begin{document}

\title{Pseudomode approach to Fano effect in dissipative cavity quantum electrodynamics}

\author{Kazuki Kobayashi}
\affiliation{Graduate School of Science and Technology, Shizuoka University, Shizuoka 422-8529, Japan}

\author{Tatsuro Yuge}
\affiliation{Graduate School of Science and Technology, Shizuoka University, Shizuoka 422-8529, Japan}
\affiliation{Department of Physics, Shizuoka University, Shizuoka 422-8529, Japan}

\begin{abstract}
  We study the Fano effect in dissipative cavity quantum electrodynamics (QED), which originates from the interference between the emitter's direct radiation and that mediated by a cavity mode. Starting from a two-level system coupled to a structured reservoir, we show that a quantum master equation previously derived within the Born-Markov approximation can be rederived by introducing a single auxiliary mode through the pseudomode approach. We identify the corresponding spectral function of the system--environment interaction and show that it consists of a constant contribution and a non-Lorentzian contribution, whose interplay gives rise to a spectral profile of the Fano form. The constant contribution represents a Markovian background and is essential for obtaining a Lindblad master equation. Furthermore, by applying Fano diagonalization to an atom--cavity system coupled to common and independent reservoirs, we independently derive the same spectral function and clarify its physical origin. Our results provide a unified description of the Fano effect in single-mode cavity QED systems and reveal its non-Markovian origin encoded in the spectral function of the structured reservoir.
\end{abstract}

\maketitle


\section{Introduction}\label{sec:introduction}

The Fano effect arises from the interference between two transition pathways: a direct transition from a discrete state to an environment (continuum) and an indirect transition to the same environment mediated by another discrete state \cite{fano1961effects}.
This interference leads to an asymmetric line shape in the system's spectral response.
The Fano effect has been observed in a wide variety of physical systems, ranging from mesoscopic physics \cite{miroshnichenko2010fano} to quantum optics \cite{fleischhauer2005electromagnetically, limonov2017fano}.

From a theoretical perspective, Fano's original work \cite{fano1961effects} also introduced a method for constructing eigenstates that include both system and environmental degrees of freedom.
This method, known as Fano diagonalization, has attracted renewed interest in the study of non-Markovian dynamics in open quantum systems \cite{dalton2001theory}.
In this context, non-Markovian dynamics arise from memory effects induced by structured reservoirs and their interaction with the system, a situation frequently encountered in nanophotonics and quantum optics \cite{lambropoulos2000fundamental}.
Revisiting the Fano effect within the framework of open quantum systems is therefore important for deepening our fundamental understanding and for exploring further applications of non-Markovian effects \cite{finkelstein2018ubiquity}.

In recent years, several quantum master equations (QMEs) that incorporate the Fano effect have been proposed for cavity quantum electrodynamics (QED) systems \cite{franke2019quantization, denning2019quantum, medina2021few, yamaguchi2021theory}.
Some of these approaches are based on Markovian embedding methods, such as the quasinormal modes approach \cite{franke2019quantization} and the few-mode mapping \cite{denning2019quantum,medina2021few},
in which an emitter is treated as the system of interest, while cavity modes appear as auxiliary degrees of freedom introduced for the embedding.
Another QME relies on the Born-Markov approximation, where both the emitter and a cavity mode are treated as the system of interest interacting with a common environment \cite{yamaguchi2021theory}.

Although these studies demonstrate that the Fano effect can be incorporated into QMEs for cavity QED systems, the resulting descriptions differ among these approaches.
In particular, QMEs based on the Markovian embedding methods describe the interference between cavity modes, whereas the QME derived within the Born-Markov approximation captures the interference between the emitter's direct radiation and the radiation mediated by a cavity mode.
As a result, it remains unclear whether the Fano effect can be treated in a unified manner in cavity QED systems.
Moreover, within the Born-Markov approach, the spectral structure of the environment responsible for the Fano-type interference is not explicitly identified.
Consequently, it remains unclear what kind of structured reservoir can physically realize the corresponding open-system dynamics.
In this study, we investigate the spectral origin of the Fano-type QME derived in Ref.~\cite{yamaguchi2021theory} by employing the pseudomode approach \cite{garraway1997nonperturbative,breuer2002theory,pleasance2020generalized,pleasance2021pseudomode}, which is one of the Markovian embedding methods.
The pseudomode formulation allows us to identify the spectral function associated with the QME explicitly and thereby establish a direct connection between the effective Lindblad dynamics and its underlying reservoir structure.

We show that the spectral function consists of a constant contribution and a non-Lorentzian contribution, and that its overall form coincides with the Fano formula.
The constant contribution represents a Markovian background, whereas the non-Lorentzian contribution originates from a structured environment and gives rise to a non-Markovian memory kernel.
The Fano effect can be understood as interference between Markovian and non-Markovian contributions encoded in the spectral function.
Our results therefore provide a physical interpretation of the Fano-type QME in terms of a structured reservoir.

Furthermore, using Fano diagonalization for a common-environment atom--cavity system, we independently derive the same spectral function.
This agreement establishes the physical relevance of the spectral structure obtained within the pseudomode framework and elucidates the non-Markovian origin of the Fano effect encoded in the spectral function of a common-environment setup.

This paper is organized as follows.
In Sec.~\ref{sec:setup}, we present the original setup and preliminaries for the pseudomode approach \cite{garraway1997nonperturbative,breuer2002theory}.
In Sec.~\ref{subsec:derivation_QME}, we derive the QME in Ref.~\cite{yamaguchi2021theory} using the pseudomode approach, where we introduce the cavity mode as an auxiliary system.
In Sec.~\ref{subsec:spectral_func}, we specify the spectral function for the present problem and show that the Fano effect is incorporated in it.
In Sec.~\ref{subsec:constant_term}, we discuss roles of the constant term in the spectral function.
We also derive the spectral function in another method relying on Fano diagonalization for the perfect interference case (Sec.~\ref{subsec:perfect_case}) and the imperfect interference case (Sec.~\ref{subsec:imperfect_case}).
Section~\ref{sec:summary} provides concluding remarks.
We set $\hbar=1$ throughout the present paper.


\section{Setup and Preliminaries}\label{sec:setup}

\begin{figure}[tb]
  \includegraphics[width=\linewidth]{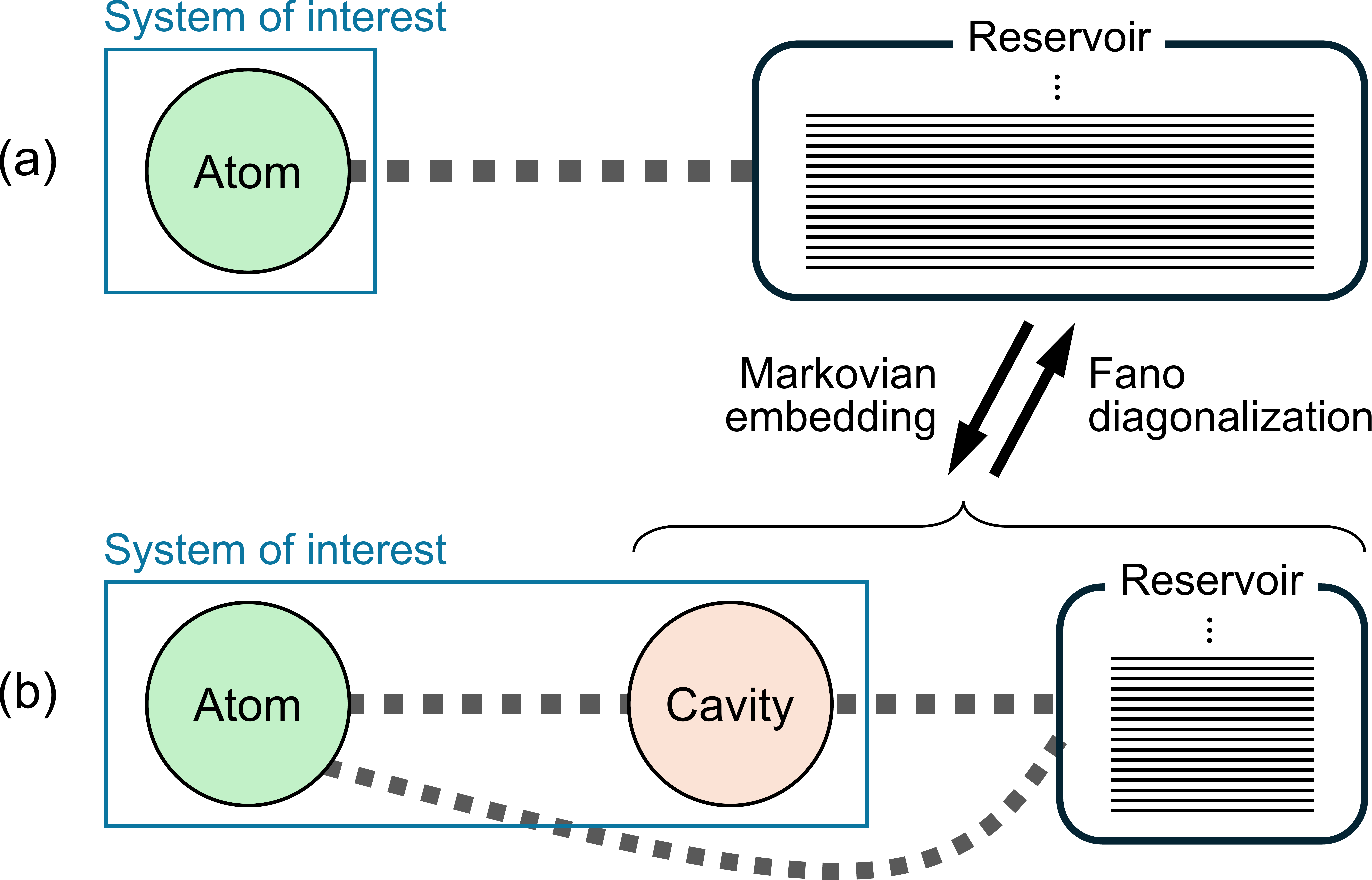}
  \caption{\label{fig:setup}
    Setups of the present study.
    (a) Original setup for the pseudomode approach (Secs.~\ref{sec:setup} and \ref{sec:pseudomode_approach}), which consists of an atom (two-level system) interacting with a reservoir (continuum of states). In this setup, the atom is the system of interest.
    (b) Setup after the Markovian embedding, which consists of the atom and a cavity (auxiliary system) interacting with a reservoir. In this setup, the atom and the cavity are the system of interest.
    The dotted lines represent interactions between the systems.
    Even after the Markovian embedding, we retain the atom--reservoir interaction that induces a Markovian dissipation for the atom.
    In Sec.~\ref{sec:Fano_diagonalization}, we perform the inverse transformation, from (b) to (a), using Fano diagonalization.
  }
\end{figure}

At the initial stage of the pseudomode approach in this study, we consider a two-level system (atom) interacting with a reservoir of harmonic oscillators, as dipicted in Fig.~\ref{fig:setup}(a).
The cavity mode is introduced at a later stage as an auxiliary system via Markovian embedding.
The total Hamiltonian
$\hat{H} = \hat{H}_{\mathrm{A}} + \hat{H}_{\mathrm{R}} + \hat{H}_{\mathrm{I}}$
is composed of the atom's free Hamiltonian $\hat{H}_{\mathrm{A}}$, reservoir's free Hamiltonian $\hat{H}_{\mathrm{R}}$, and interaction Hamiltonian $\hat{H}_{\mathrm{I}}$:
\begin{align}
  \label{H_system_setup}
  \hat{H}_{\mathrm{A}} & = \omega_{\mathrm{A}} \hat{\sigma}^{\dagger} \hat{\sigma},
  \\
  \label{H_reservoir_setup}
  \hat{H}_{\mathrm{R}} & = \sum_{k} \omega_{k} \hat{b}_{k}^{\dagger} \hat{b}_{k},
  \\
  \label{H_interaction_setup}
  \hat{H}_{\mathrm{I}} & = \sum_{k} \bigl( g_{k} \hat{\sigma}^{\dagger} \hat{b}_{k} + g_{k}^{*} \hat{\sigma} \hat{b}_{k}^{\dagger} \bigr).
\end{align}
Here $\omega_{\mathrm{A}}$ and $\omega_{k}$ are the transition frequency of the atom and the reservoir frequency for mode $k$, respectively, and $g_{k}$ is the coupling constant between the atom and the mode $k$.
The operators are defined as follows: $\hat{\sigma}^{\dagger}$ and $\hat{\sigma}$ are the atom's raising and lowering operators,
and $\hat{b}_{k}$ and $\hat{b}_{k}^{\dagger}$ are the reservoir's annihilation and creation operators for mode $k$, satisfying $[ \hat{b}_{k}, \hat{b}_{k'}^{\dagger} ] = \delta_{k,k'}$.

In the present paper, we assume, for simplicity, that the total system contains at most one excitation.
This is a standard assumption in the pseudomode approach \cite{garraway1997nonperturbative,breuer2002theory},
although generalized formulations of Markovian embedding allowing for multiple excitations have been developed recently \cite{lambert2019modelling,pleasance2020generalized,pleasance2021pseudomode,menczel2024nonhermitian,PhysRevB.110.195148,huang2026coupled}.
Thus the initial state of the total system has the following form:
\begin{align}
  \ket{\Psi(0)} = c_{0}(0)\ket{\Psi_{0}} + c_{1}(0)\ket{\Psi_{1}} + \sum_{k} c_{k}(0)\ket{\Psi_{k}},
  \label{initial_state_tot}
\end{align}
where
$\ket{\Psi_{0}} = \ket{0}_{\mathrm{A}} \ket{0}_{\mathrm{R}}$,
$\ket{\Psi_{1}} = \ket{1}_{\mathrm{A}} \ket{0}_{\mathrm{R}}$,
and $\ket{\Psi_{k}} = \ket{0}_{\mathrm{A}} \ket{k}_{\mathrm{R}}$.
Here, $\ket{0}_{\mathrm{A}}$ and $\ket{1}_{\mathrm{A}} = \hat{\sigma}^{\dagger} \ket{0}_{\mathrm{A}}$ denote the ground and excited states of the atom, respectively.
$\ket{0}_{\mathrm{R}}$ and $\ket{k}_{\mathrm{R}} = \hat{b}_{k}^{\dagger}\ket{0}_{\mathrm{R}}$ denote the vacuum and singly-excited states of the reservoir, respectively.
In the following, we further assume that the reservoir is initially in the vacuum state so that $c_{k}(0) = 0$.

In this setup, the total excitation number $\hat{N} = \hat{\sigma}^{\dagger} \hat{\sigma} + \sum_{k}\hat{b}_{k}^{\dagger}\hat{b}_{k}$ is a conserved quantity, since it commutes with $\hat{H}$.
As a result, the state $\ket{\Psi(t)}$ of the total system at time $t$ has the same form as that of the initial state:
\begin{align}
  \label{def_total_state_vector_Psi_t}
  \ket{\Psi(t)} = c_{0}(t)\ket{\Psi_{0}} + c_{1}(t)\ket{\Psi_{1}} + \sum_{k}c_{k}(t)\ket{\Psi_{k}}.
\end{align}
By substituting Eq.~\eqref{def_total_state_vector_Psi_t} into the Schr\"odinger equation,
\begin{align}
  \label{sch_equation_total_system}
  \frac{d}{dt}\ket{\Psi(t)} = -i\hat{H}\ket{\Psi(t)},
\end{align}
we obtain the time-evolution equations for the amplitudes:
\begin{align}
  \label{derivative_of_c_0_t_setup}
   & \frac{d}{dt}c_{0}(t) = 0,
  \\
  \label{derivative_of_c_1_t_setup}
   & \frac{d}{dt}c_{1}(t) = -i \omega_{\mathrm{A}}c_{1}(t) -i \sum_{k} g_{k}c_{k}(t),
  \\
  \label{derivative_of_b_k_t_setup}
   & \frac{d}{dt}c_{k}(t) = -i \omega_{k}c_{k}(t) -i g_{k}^{*}c_{1}(t).
\end{align}
From Eq.~\eqref{derivative_of_c_0_t_setup}, we have $c_{0}(t) = c_{0}(0) = c_{0}$.

To obtain the reduced dynamics of the atom alone, we eliminate the reservoir degrees of freedom
by integrating Eq.~\eqref{derivative_of_b_k_t_setup} under the initial condition $c_{k}(0) = 0$
and substituting the result into Eq.~\eqref{derivative_of_c_1_t_setup}.
The resulting exact reduced dynamical equation is
\begin{align}
  \label{eq_c_1_t_sch_pic_not_pseudo}
  \frac{d}{dt}c_{1}(t)
  = - i \omega_{\mathrm{A}} c_{1}(t) - \int_{0}^{t}dt' \, F(t - t') c_{1}(t'),
\end{align}
where the memory kernel $F(\tau)$ is given by
\begin{align}
  \label{def_memory_kernel_F}
  F(\tau) = \int d\omega \, J(\omega) e^{-i\omega\tau}.
\end{align}
Here $J(\omega)$ is the spectral function defined by
\begin{align}
  J(\omega) = \sum_{k}|g_{k}|^{2}\delta(\omega - \omega_{k}),
\end{align}
which characterizes the structure of the interaction between the atom and the reservoir.
Thus, the reduced dynamics of the atom in this setup [Fig.~\ref{fig:setup}(a)] is governed by the non-Markovian equation~\eqref{eq_c_1_t_sch_pic_not_pseudo}, where the non-Markovian memory effect is encoded in the spectral function $J(\omega)$ through the kernel $F(\tau)$.
In the next section, we map this non-Markovian atomic dynamics onto a Markovian dynamics of an enlarged system of the atom and a pseudomode,
and show that the Fano effect is also encoded in the spectral function $J(\omega)$.


\section{Pseudomode approach}\label{sec:pseudomode_approach}

The present paper has two main objectives.
First, we show that the QME derived in Ref.~\cite{yamaguchi2021theory} (hereafter referred to as Yamaguchi's QME), previously obtained within the Born-Markov approximation, can be rederived using the pseudomode approach.
Second, we identify the concrete form of $J(\omega)$ responsible for the Fano interference between the direct atomic radiation and the radiation mediated by the cavity mode.

Before proceeding to the derivation, let us first present Yamaguchi's QME:
\begin{align}
  \label{goal_of_the_QME}
  \frac{d}{dt} \hat{\rho}(t)
   & = - i \bigl[ \hat{H}_{0} , \hat{\rho}(t) \bigr]
  + \mathcal{D}_{\mathrm{A}}(\hat{\rho}(t))
  + \mathcal{D}_{\mathrm{C}}(\hat{\rho}(t))
  + \mathcal{D}_{\mathrm{F}}(\hat{\rho}(t)).
\end{align}
$\hat{H}_{0}$ is the Hamiltonian of the coupled atom--cavity system:
\begin{align}
  \label{goal_of_the_QME_Hamiltonian_part}
  \hat{H}_{0} = \omega_{\mathrm{A}} \hat{\sigma}^{\dagger} \hat{\sigma}
  + \omega_{\mathrm{C}} \hat{a}^{\dagger} \hat{a}
  + g \hat{\sigma}^{\dagger} \hat{a} + g^* \hat{\sigma} \hat{a}^{\dagger},
\end{align}
where $\hat{\sigma}$ and  $\omega_{\mathrm{A}}$ are the same as those in Eq.~\eqref{H_system_setup},
$\hat{a}$ and $\hat{a}^\dag$ are the bosonic annihilation and creation operators of the cavity system,
$\omega_{\mathrm{C}}$ is the cavity frequency, and $g = |g| e^{i\phi}$ is the complex-valued coupling constant between the atom and the cavity.
The dissipators are given by
\begin{align}
  \label{dissipator_atom}
  \mathcal{D}_{\mathrm{A}}(\hat{\rho})
   & = \gamma \Big( \hat{\sigma} \hat{\rho} \hat{\sigma}^{\dagger} - \frac{1}{2} \{ \hat{\sigma}^{\dagger} \hat{\sigma} , \hat{\rho} \} \Big),
  \\
  \mathcal{D}_{\mathrm{C}}(\hat{\rho})
   & = \kappa \Big( \hat{a} \hat{\rho} \hat{a}^{\dagger} - \frac{1}{2} \{ \hat{a}^{\dagger} \hat{a} , \hat{\rho} \} \Big),
  \\
  \label{dissipator_Fano}
  \mathcal{D}_{\mathrm{F}}(\hat{\rho})
   & = \gamma_{\mathrm{F}} \Big( \hat{a} \hat{\rho} \hat{\sigma}^{\dagger} - \frac{1}{2} \{ \hat{\sigma}^{\dagger} \hat{a} , \hat{\rho} \}
  \Big) + \mathrm{h.c.},
\end{align}
where $\gamma$ and $\kappa$ are the loss rates of the atom and the cavity, respectively.
$\mathcal{D}_{\mathrm{F}}$ describes the Fano interference between the radiation directly form the atom and that mediated by the cavity,
and $\gamma_{\mathrm{F}} = \sqrt{\eta \gamma \kappa} e^{i(\theta_{\mathrm{A}} - \theta_{\mathrm{C}})}$ is the complex-valued rate associated with the Fano effect.
Here, $\eta$ represents the strength of the Fano effect ($0 \leq \eta \leq 1$),
controlled by the overlap between the spatial radiation patterns of the atom and the cavity \cite{yamaguchi2021theory},
and $\theta_{\mathrm{A}} - \theta_{\mathrm{C}}$ is the relative phase between the atom--reservoir and cavity--reservoir couplings.

\subsection{Derivation of the QME}\label{subsec:derivation_QME}

We now derive Yamaguchi's QME~\eqref{goal_of_the_QME} by applying the pseudomode approach to the original setup [Fig.~\ref{fig:setup}(a)] introduced in Sec.~\ref{sec:setup}.
In the pseudomode approach, the original non-Markovian dynamics is mapped onto an equivalent Markovian dynamics in the setup in Fig.~\ref{fig:setup}(b), where the system of interest is enlarged (the atom and an auxiliary cavity mode).
These two descriptions are equivalent in the sense that they give the same reduced dynamics for the atom.

The outline of the derivation is as follows (details are given below).
First, we impose several assumptions on the spectral function $J(\omega)$, which allow us to determine the form of the memory kernel $F(\tau)$.
We then introduce the pseudomode amplitude and rewrite the non-Markovian equation~\eqref{eq_c_1_t_sch_pic_not_pseudo} for the atom alone into coupled memoryless equations for the enlarged system composed of the atom and the pseudomode.
Next, we construct a non-Hermitian effective Hamiltonian for the enlarged system whose Schr\"odinger equation reproduces these coupled equations.
Finally, by incorporating the quantum-jump contribution, we arrive at a Markovian QME for the enlarged system.

We first consider the conditions on the spectral function $J(\omega)$.
In our original setup given in Eqs.~\eqref{H_system_setup}--\eqref{H_interaction_setup}, the atom is the system of interest.
Consequently, the atomic dissipator $\mathcal{D}_{\mathrm{A}}$ in Eq.~\eqref{dissipator_atom} represents a genuinely Markovian decay for the system.
Therefore, to derive $\mathcal{D}_{\mathrm{A}}$ within the pseudomode approach,
$J(\omega)$ must contain a constant term $J_0$ that encodes the memoryless character of this dissipation:
\begin{align}
  \label{spectral_func_1}
  J(\omega) = J_0 + f(\omega).
\end{align}
The frequency-dependent part $f(\omega)$ represents the non-Markovian effect.
We furthermore impose the following conditions on $f(\omega)$:
\begin{enumerate}[label=(\alph*)]
  \item \label{cond:pole} $f(z)$ has only a single simple pole $z_{1}$ in the lower half of the complex plane.
  \item \label{cond:residue} The corresponding residue $r_{1} = \mathrm{Res}(f(z); z_{1})$ is a complex number.
  \item \label{cond:real} $f(z)$ is real-valued on the real axis.
\end{enumerate}
Condition~\ref{cond:pole} ensures that only a single auxiliary mode is introduced in the pseudomode approach, which implies that the resulting QME involves a single-mode cavity.
Condition~\ref{cond:residue} is required to incorporate the Fano effect, as will become clear later.
Additionally, we assume that $f(z)$ is meromorphic in the lower-half complex plane and vanishes sufficiently rapidly as $|z| \to \infty$ in the lower-half plane.
The explicit form of $f(\omega)$ will be specified in Sec.~\ref{subsec:spectral_func}.

Under these assumptions on $J(\omega)$, using the residue theorem we can evaluate the memory kernel in Eq.~\eqref{def_memory_kernel_F} as
\begin{align}
  F(\tau) = 2 \pi J_0 \delta(\tau) - 2 \pi i r_{1} e^{-i z_{1} \tau}.
\end{align}
By substituting this kernel into Eq.~\eqref{eq_c_1_t_sch_pic_not_pseudo}, we obtain
\begin{align}
  \frac{d}{dt} c_{1}(t)
   & = - i ( \omega_{\mathrm{A}} - i \pi J_{0} ) c_{1}(t)
  \notag
  \\
   & \quad + 2 \pi i r_{1} \int_{0}^{t} dt' \, e^{ -i z_{1} (t - t') } c_{1}(t').
  \label{eq_c_1_t_sch_pic_not_pseudo_2_gamma}
\end{align}

We now introduce a pseudomode amplitude $b_{1}(t)$ as follows.
We factorize the residue as $2 \pi i r_{1} = - \tilde{g}_- \tilde{g}_+^*$ with two complex numbers, $\tilde{g}_-$ and $\tilde{g}_+$.
We then rewrite the second term in Eq.~\eqref{eq_c_1_t_sch_pic_not_pseudo_2_gamma} as $-i \tilde{g}_- b_{1}(t)$,
where the amplitude is defined by
\begin{align}
  \label{def_b_1_t_1}
  b_{1}(t) = - i \tilde{g}_+^* \int_{0}^{t} dt' \, e^{ -i z_{1} (t - t') } c_{1}(t').
\end{align}
Then, Eq.~\eqref{eq_c_1_t_sch_pic_not_pseudo_2_gamma} becomes
\begin{align}
  \label{eq_derivative_c_1_t_in_sch_pic_pseudo}
  \frac{d}{dt} c_{1}(t)
  = - i ( \omega_{\mathrm{A}} - i \pi J_{0} ) c_{1}(t) - i \tilde{g}_- b_{1}(t),
\end{align}
and the time derivative of $b_{1}(t)$ yields
\begin{align}
  \label{eq_derivative_b_1_t_in_sch_pic_pseudo}
  \frac{d}{dt} b_{1}(t) = -i z_{1} b_{1}(t) - i \tilde{g}_+^* c_{1}(t).
\end{align}
Thus, the differential equation~\eqref{eq_c_1_t_sch_pic_not_pseudo} for $c_{1}(t)$ containing the memory kernel is transformed into the coupled memoryless equations~\eqref{eq_derivative_c_1_t_in_sch_pic_pseudo} and \eqref{eq_derivative_b_1_t_in_sch_pic_pseudo}.
We note that in the standard pseudomode approach \cite{garraway1997nonperturbative}, the residue is factorized as $2 \pi i r_1 = - \sqrt{-2 \pi i r_1} \times \sqrt{-2 \pi i r_1}$, whereas in the present study we factorize it as $2 \pi i r_{1} = - \tilde{g}_- \tilde{g}_+^*$.
This alternative factorization is essential for incorporating the phase factors in the atom--pseudomode coupling and the rate associated with the Fano effect.

Next, we introduce an auxiliary bosonic system (pseudomode) and a non-Hermitian effective Hamiltonian for the enlarged system composed of the atom and the pseudomode,
such that the corresponding Schr\"odinger equation reproduces the coupled equations~\eqref{eq_derivative_c_1_t_in_sch_pic_pseudo} and \eqref{eq_derivative_b_1_t_in_sch_pic_pseudo}.
Since the auxiliary mode is bosonic, we can interpret it as a single-mode cavity system
and therefore use the same symbols as those in Yamaguchi's QME~\eqref{goal_of_the_QME}, $\hat{a}$ and $\hat{a}^{\dagger}$, for its annihilation and creation operators.
Its vacuum state is denoted by $\ket{0}_{\mathrm{C}}$, and the singly-excited state by $\ket{1}_{\mathrm{C}} = \hat{a}^{\dagger} \ket{0}_{\mathrm{C}}$.

We consider the enlarged system as the system of interest [Fig.~\ref{fig:setup}(b)].
As in Eqs.~\eqref{initial_state_tot} and \eqref{def_total_state_vector_Psi_t}, we restrict the analysis to the subspace of states that include at most one excitation and express the state vector of the enlarged system as
\begin{align}
  \label{def_state_vector_of_total_system_in_Sch_pic_pseudomode}
  \ket{\tilde{\Psi}(t)}
   & = c_{0} \ket{\tilde{\Psi}_{0}}
  + c_{1}(t) \ket{\tilde{\Psi}_{1}}
  + b_{1}(t) \ket{\tilde{\Phi}_{1}},
\end{align}
where $\ket{\tilde{\Psi}_{0}} = \ket{0}_{\mathrm{A}} \ket{0}_{\mathrm{C}}$,
$\ket{\tilde{\Psi}_{1}} = \ket{1}_{\mathrm{A}} \ket{0}_{\mathrm{C}}$,
and $\ket{\tilde{\Phi}_{1}} = \ket{0}_{\mathrm{A}} \ket{1}_{\mathrm{C}}$.
Equation~\eqref{def_state_vector_of_total_system_in_Sch_pic_pseudomode} is consistent with the fact that $b_{1}(t)$ is the pseudomode amplitude.
We note that $\ket{\tilde{\Psi}(t)}$ is not normalized
since $c_{1}(t)$ and $b_{1}(t)$ evolve according to Eqs.~\eqref{eq_derivative_c_1_t_in_sch_pic_pseudo} and \eqref{eq_derivative_b_1_t_in_sch_pic_pseudo}, 
which represent dissipative dynamics.
The time derivative of $\ket{\tilde{\Psi}(t)}$ can then be written in the form of a Schr\"odinger equation:
\begin{align}
  \label{sch_eq_at_total_system_in_sch_pic_pseudomode}
  \frac{d}{dt} \ket{\tilde{\Psi}(t)} = - i \hat{H}_{\text{eff}} \ket{\tilde{\Psi}(t)},
\end{align}
where the effective Hamiltonian $\hat{H}_{\text{eff}}$ is given by
\begin{align}
   & \hat{H}_{\text{eff}}
  = (\omega_{\mathrm{A}} - i \pi J_{0} ) \hat{\sigma}^{\dagger} \hat{\sigma}
  + z_{1} \hat{a}^{\dagger} \hat{a}
  + \tilde{g}_- \hat{\sigma}^{\dagger} \hat{a} + \tilde{g}_+^* \hat{\sigma} \hat{a}^{\dagger}
  \notag
  \\
   & = \hat{H}_{\mathrm{AC}}
  - i \bigl( \pi J_{0} \hat{\sigma}^{\dagger} \hat{\sigma}
  - \Im z_{1} \hat{a}^{\dagger} \hat{a}
  + \nu \hat{\sigma}^{\dagger} \hat{a} + \nu^* \hat{\sigma} \hat{a}^{\dagger} \bigr),
  \label{def_effective_Hamiltonian_pseudomode}
  \\
  \label{def_atom_and_auxiliary_system_Hamiltonian}
   & \hat{H}_{\mathrm{AC}}
  = \omega_{\mathrm{A}} \hat{\sigma}^{\dagger} \hat{\sigma} + \Re z_{1} \hat{a}^{\dagger} \hat{a}
  + \mu \hat{\sigma}^{\dagger} \hat{a} + \mu^* \hat{\sigma} \hat{a}^{\dagger}.
\end{align}
In the above equations,
we have introduced two complex parameters, $\mu$ and $\nu$, satisfying $\tilde{g}_\pm = \mu \pm i\nu$.
The derivation of $\hat{H}_{\text{eff}}$ [the first line of Eq.~\eqref{def_effective_Hamiltonian_pseudomode}] is given in Appendix~\ref{appendix:derivation_of_H_eff}.

This result shows that the enlarged system [the system of interest in Fig.~\ref{fig:setup}(b)] governed by the effective Hamiltonian $\hat{H}_{\text{eff}}$ in Eq.~\eqref{def_effective_Hamiltonian_pseudomode} provides an equivalent Markovian description of the original atom--reservoir setup in Eqs.~\eqref{H_system_setup}--\eqref{H_interaction_setup}, or equivalently, the non-Markovian equation~\eqref{eq_c_1_t_sch_pic_not_pseudo}, with the spectral function $J(\omega)$ under Conditions~\ref{cond:pole} and \ref{cond:residue}.
In particular, eliminating the pseudomode degree of freedom from Eqs.~\eqref{eq_derivative_c_1_t_in_sch_pic_pseudo} and \eqref{eq_derivative_b_1_t_in_sch_pic_pseudo} reproduces the non-Markovian equation~\eqref{eq_c_1_t_sch_pic_not_pseudo} for the atom alone.

We note that $\hat{H}_{\text{eff}}$ is non-Hermitian because $z_{1}$ and $\tilde{g}_\pm$ are complex, reflecting the dissipative nature of the effective dynamics.
We also note that $\hat{H}_{\text{eff}}$ commutes with the excitation number $\hat{\sigma}^\dag \hat{\sigma} + \hat{a}^{\dagger} \hat{a}$ in the enlarged system.

To derive the QME for the enlarged system, we introduce the following Hermitian operator:
\begin{align}
  \label{def_rho_nj_t}
  \hat{\rho}_{\mathrm{nj}}(t) = \ketbra{\tilde{\Psi}(t)}.
\end{align}
Differentiating this operator and using Eq.~\eqref{sch_eq_at_total_system_in_sch_pic_pseudomode}, we obtain
\begin{align}
  \label{eq_derivative_rho_nj_t}
  \frac{d}{dt} \hat{\rho}_{\mathrm{nj}} (t)
   & = - i \Big( \hat{H}_{\text{eff}} \hat{\rho}_{\mathrm{nj}} (t) - \hat{\rho}_{\mathrm{nj}}(t) \hat{H}_{\text{eff}}^{\dagger} \Big).
\end{align}
Since $\hat{H}_{\text{eff}}$ is non-Hermitian, this equation does not preserve the trace of $\hat{\rho}_{\mathrm{nj}}(t)$.
To restore trace preservation, the pseudomode approach requires an additional contribution associated with quantum jump, described by the following Hermitian operator $\hat{\rho}_{\mathrm{j}}(t)$:
\begin{align}
  \label{def_rho_j_t}
  \hat{\rho}_{\mathrm{j}} (t) = \Pi_{\mathrm{j}}(t) \ketbra{\tilde{\Psi}_{0}},
\end{align}
where $\Pi_{\mathrm{j}}(t) = 1 - |c_{0}|^{2} - |c_{1}(t)|^{2} - |b_{1}(t)|^{2}$
is the quantum-jump contribution to the probability of the atom being in the ground state.
The time derivative of this operator is given in Appendix~\ref{sec:appendix}.
We then obtain the trace-preserving dynamical equation for
the density matrix
$\hat{\rho}(t) = \hat{\rho}_{\mathrm{j}}(t) + \hat{\rho}_{\mathrm{nj}}(t)$
of the enlarged atom--pseudomode system as
\begin{align}
  \label{the_QME_via_Pseudomode_approach}
  \frac{d}{dt} \hat{\rho}(t)
   & = - i \bigl[ \hat{H}_{\mathrm{AC}} , \hat{\rho}(t) \bigr]
  + 2 \pi J_0 \Big( \hat{\sigma} \hat{\rho}(t) \hat{\sigma}^{\dagger} - \frac{1}{2} \{ \hat{\sigma}^{\dagger} \hat{\sigma} , \hat{\rho}(t) \} \Big)
  \notag
  \\
   & \quad - 2 \Im z_{1} \Big( \hat{a} \hat{\rho}(t) \hat{a}^{\dagger} - \frac{1}{2} \{ \hat{a}^{\dagger} \hat{a} , \hat{\rho}(t) \} \Big)
  \notag
  \\
   & \quad + 2 \nu \Big( \hat{a} \hat{\rho}(t) \hat{\sigma}^{\dagger} - \frac{1}{2} \{ \hat{\sigma}^{\dagger} \hat{a} , \hat{\rho}(t) \} \Big) + \mathrm{h.c.}
\end{align}
Equation~\eqref{the_QME_via_Pseudomode_approach} is the QME that describes the dynamics of the enlarged system in Fig.~\ref{fig:setup}(b), rather than the original atom-only system in Fig.~\ref{fig:setup}(a).
However, the two descriptions are equivalent at the level of the reduced atomic dynamics: tracing out the pseudomode degree of freedom in Eq.~\eqref{the_QME_via_Pseudomode_approach} yields the same atomic dynamics as that generated by the non-Markovian equation~\eqref{eq_c_1_t_sch_pic_not_pseudo}.
Therefore, Eq.~\eqref{the_QME_via_Pseudomode_approach} can be regarded as a Markovian embedding of the original atom--reservoir setup.

The QME \eqref{the_QME_via_Pseudomode_approach} with Eq.~\eqref{def_atom_and_auxiliary_system_Hamiltonian} has the same form as that of Yamaguchi's QME~\eqref{goal_of_the_QME} with Eqs.~\eqref{goal_of_the_QME_Hamiltonian_part}--\eqref{dissipator_Fano}.
Therefore, by comparing the coefficients, we obtain the following correspondence:
\begin{align}
  \label{correspondence_alpha}
   & J_0 = \frac{\gamma}{2 \pi},
  \\
   & \Re z_{1} = \omega_{\mathrm{C}}, \quad
  \Im z_{1} = - \frac{\kappa}{2},
  \\
  \label{correspondence_g1}
   & \mu = g, \quad
  \nu = \frac{\gamma_{\mathrm{F}}}{2}.
\end{align}
We note that $J_0$, $\Im z_1$, and $\nu$ are related through $\gamma_{\mathrm{F}} = \sqrt{\eta \gamma \kappa} e^{i(\theta_{\mathrm{A}} - \theta_{\mathrm{C}})}$ ($0 \leq \eta \leq 1$).
Furthermore, from $\tilde{g}_\pm = \mu \pm i\nu$ and $2 \pi i r_1 = -\tilde{g}_- \tilde{g}_+^*$, we have
\begin{align}
  \label{g_pm}
  \tilde{g}_\pm
   & = g \pm i \frac{\gamma_{\mathrm{F}}}{2}
  = |g| e^{i\phi} \pm \frac{i}{2} \sqrt{\eta \gamma \kappa} e^{i(\theta_{\mathrm{A}} - \theta_{\mathrm{C}})},
  \\
  \label{result_of_residue}
  r_{1}
   & = \frac{i}{2 \pi} \Bigl( g - i\frac{\gamma_{\mathrm{F}}}{2} \Bigr) \Bigl(g^* - i\frac{\gamma_{\mathrm{F}}^*}{2} \Bigr)
  \notag
  \\
   & = \frac{i}{2 \pi} \Bigl[ |g|^2 - \frac{\eta \gamma \kappa}{4} - i |g| \sqrt{\eta \gamma \kappa} \cos(\Delta \phi) \Bigr],
\end{align}
where $\Delta \phi = \phi - \theta_{\mathrm{A}} + \theta_{\mathrm{C}}$.

Equation~\eqref{g_pm} clarifies the physical meaning of the effective atom--pseudomode coupling $\tilde{g}_\pm$ appearing in Eq.~\eqref{def_effective_Hamiltonian_pseudomode}.
The coupling consists of two distinct contributions.
The term proportional to $g$ represents the direct coherent interaction between the atom and the cavity mode (pseudomode), and therefore appears in the Hermitian part $\hat{H}_{\mathrm{AC}}$ of the effective Hamiltonian $\hat H_{\mathrm{eff}}$.
In contrast, the term proportional to $\gamma_{\mathrm F}$ represents the interaction with the common reservoir shared by the atom and the cavity mode.
This reservoir-mediated contribution appears in the non-Hermitian part of $\hat H_{\mathrm{eff}}$ and represents a dissipative coupling through the environment, which induces the Fano interference.
The dependence $\gamma_{\mathrm F} \propto \sqrt{\eta\gamma\kappa}$ reflects the reservoir-mediated origin of this coupling, since its strength is determined jointly by the atomic decay rate $\gamma$, the cavity decay rate $\kappa$, and the strength of the Fano effect $\eta$.
Therefore, the effective coupling $\tilde{g}_\pm$ in Eq.~\eqref{def_effective_Hamiltonian_pseudomode} shows the interplay between the direct coherent process and the reservoir-mediated dissipative process underlying the Fano effect.

\subsection{Spectral function}\label{subsec:spectral_func}

We now determine the specific form of the spectral function $J(\omega)$.
Under Conditions \ref{cond:pole}--\ref{cond:real}
we can express $f(\omega)$ in terms of its pole $z_{1}$ and residue $r_{1}$ as
\begin{align}
  \label{result_f_omega_general_1}
  f(\omega)
   & = \frac{r_1}{\omega - z_1} + \frac{r_1^*}{\omega - z_1^*}
  \notag
  \\
   & = \frac{2 \bigl[ \Re r_{1} \big( \omega - \Re z_{1} \big) - \Im z_{1} \Im r_{1} \bigr]}{ ( \omega - \Re z_{1} ) ^{2} + (\Im z_{1})^{2} }.
\end{align}
We then use the correspondence in Eqs.~\eqref{correspondence_alpha}--\eqref{result_of_residue} to specify the form of $2 \pi J(\omega) = 2 \pi J_0 + 2 \pi f(\omega)$ as
\begin{align}
  \label{result_spectral_function}
   & 2 \pi J(\omega)
  \\
   & = \gamma + \frac{8 |g| \sqrt{\eta \gamma \kappa} (\omega - \omega_{\mathrm{C}}) \cos(\Delta \phi) + \kappa (4 |g|^{2} - \eta \gamma \kappa ) }{4 (\omega - \omega_{\mathrm{C}})^{2} + \kappa^{2} }.
  \notag
\end{align}
This form of the spectral function is one of the main results of the present paper.
By employing the pseudomode approach instead of the Born-Markov approximation used in Ref.~\cite{yamaguchi2021theory}, we can explicitly identify the form \eqref{result_spectral_function} of the spectral function underlying Yamaguchi's QME~\eqref{goal_of_the_QME}.
As discussed below, the Fano effect can be understood as arising from the interference between the constant background and the frequency-dependent contribution in this spectral function, thereby clarifying its non-Markovian origin.

\begin{figure}[tb]
  \includegraphics[width=.95\linewidth]{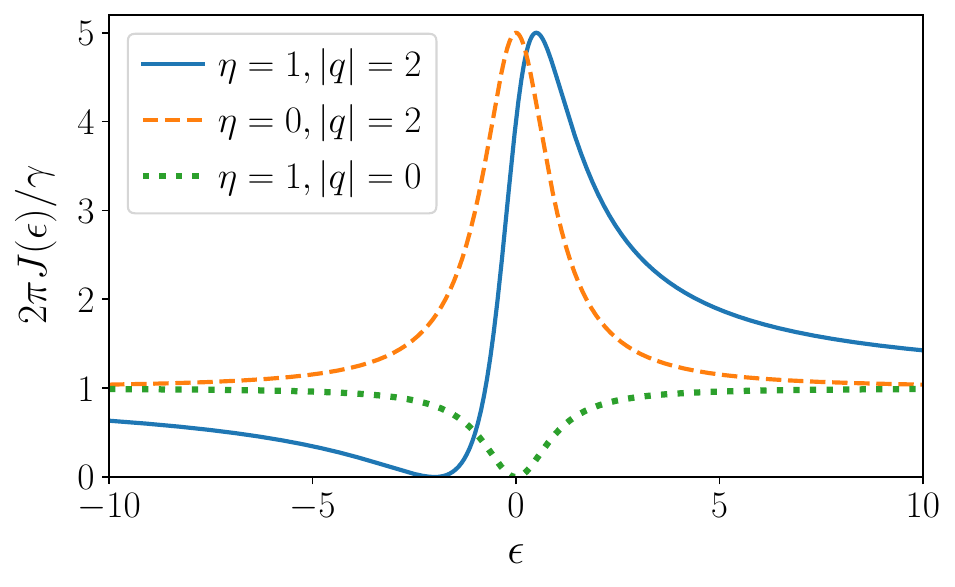}
  \caption{\label{fig:spectral_function}
    Spectral function $J(\epsilon)$, plotted against the reduced detuning $\epsilon = 2 (\omega - \omega_{\mathrm{C}}) / \kappa$.
    The solid, dashed, and dotted curves correspond to the cases of $(\eta, |q|) = (1, 2)$, $(0, 2)$, and $(1, 0)$, respectively.
    We set $\Delta \phi = 0$.
  }
\end{figure}

We can show that, in the case of $\gamma \ll \kappa$, the spectral function $2\pi J(\omega_{\mathrm{A}})$ evaluated at the atomic frequency $\omega_{\mathrm{A}}$ coincides with the transition rate $W$ of the initially excited atom.
As shown in Ref.~\cite{yamaguchi2021theory}, $W$ exhibits a Fano-type anti-resonance profile as a function of the detuning $\omega_{\mathrm{A}} - \omega_{\mathrm{C}}$.
In this sense, the Fano effect is encoded in $J(\omega)$.

To see the Fano effect directly in the spectral function, we rewrite Eq.~\eqref{result_spectral_function} as
\begin{equation}
  \label{spectral_function_e_q}
  2 \pi J(\epsilon)
  = \frac{\gamma}{ \epsilon^{2} + 1 } \big[ |\epsilon + \sqrt{\eta} q|^2 + (1 - \eta)(1 + |q|^2) \big],
\end{equation}
where $\epsilon = 2 (\omega - \omega_{\mathrm{C}}) / \kappa$ is the reduced detuning
and $q = 2 |g| e^{i \Delta \phi} / \sqrt{\gamma \kappa}$ corresponds to the Fano parameter.
For the case of the perfect Fano interference ($\eta = 1$), $J(\omega)$ reduces to
\begin{equation}
  \label{spectral_function_eta_1}
  2 \pi J(\epsilon) = \frac{\gamma |\epsilon + q|^2}{ \epsilon^{2} + 1 } \quad (\eta = 1).
\end{equation}
Since this result coincides with the (generalized) Fano formula \cite{fano1961effects, yamaguchi2021theory}, we confirm that the Fano effect is encoded in the spectral function.
The solid curve in Fig.~\ref{fig:spectral_function} is the plot of $J(\epsilon)$ for this case ($\eta = 1$ and $|q| = 2$), which exhibits an asymmetric profile with an  anti-resonance dip characteristic to the Fano effect.
In contrast, for the case of no Fano effect ($\eta = 0$), $J(\omega)$ reduces to
\begin{equation}
  2 \pi J(\epsilon) = \gamma \left[1 + \frac{|q|^2}{ \epsilon^{2} + 1 } \right] \quad (\eta = 0),
\end{equation}
which represents the Purcell effect.
The dashed curve in Fig.~\ref{fig:spectral_function} is the plot of $J(\epsilon)$ for this case ($\eta = 0$ and $|q| = 2$), which exhibits a symmetric profile.

The Fano parameter $q = 2 |g| e^{i \Delta \phi} / \sqrt{\gamma \kappa}$ contains the relative phase $\Delta \phi$ between the coupling $g = |g| e^{i \phi}$ and the rate $\gamma_{\mathrm{F}} = \sqrt{\eta \gamma \kappa} e^{i(\theta_{\mathrm{A}} - \theta_{\mathrm{C}})}$, and is therefore complex.
Such a complex-valued $q$ has been experimentally observed in quantum dot systems \cite{kobayashi2002tuning, kobayashi2003mesoscopic}.
We note that this relative phase $\Delta \phi$ cannot be eliminated by phase rotations of the operators, $\hat{\sigma} \to e^{i \phi_{\mathrm{A}}} \hat{\sigma}$ and $\hat{a} \to e^{i \phi_{\mathrm{C}}} \hat{a}$.

As another interesting case, we consider $g = 0$ ($q=0$).
In this limit, Eq.~\eqref{spectral_function_e_q} reduces to
\begin{equation}
  \label{spectral_function_g_zero}
  2 \pi J(\epsilon) = \gamma \left[1 - \frac{\eta}{ \epsilon^{2} + 1 } \right] \quad (g = 0).
\end{equation}
The dotted curve in Fig.~\ref{fig:spectral_function} is the plot of $J(\epsilon)$ for this case ($\eta = 1$ and $|q| = 0$), which exhibits symmetric profile with an  anti-resonance dip.
The first term in Eq.~\eqref{spectral_function_g_zero} can be regarded as the infinite-width limit of a Lorentzian with a positive weight. In contrast, the second term corresponds to a Lorentzian with a negative weight, which clearly represents an anti-resonance feature.
This situation is equivalent to the case discussed in Ref.~\cite{garraway1997nonperturbative}, where two auxiliary modes are present, one of which has an extremely large leakage rate while the other carries a negative weight.
It is also consistent with the discussion in Ref.~\cite{nabiev1993dynamics}.

\subsection{Roles of the constant term in the spectral function}\label{subsec:constant_term}

Here, we discuss roles of the constant term $J_0 \, (=\gamma / 2 \pi)$ in the spectral function $J(\omega)$.
As discussed above Eq.~\eqref{spectral_func_1}, we have introduced this term to include a Markovian noise acting on the atom.

First, we note that although such a constant term violates the normalization condition for the spectral function usually assumed in the pseudomode approach \cite{garraway1997nonperturbative}, it does not invalidate the derivation in Sec.~\ref{subsec:derivation_QME}.

In practical situations, this term is approximately constant within the frequency range relevant to the system dynamics and rapidly decays outside it.
Therefore, in the parameter regime considered in this work, where the atom--reservoir and atom--cavity interactions are well described within the rotating-wave approximation and the dynamics is mainly governed by the spectral structure around the relevant system frequencies, $\omega_{\mathrm{A}}$ and $\omega_{\mathrm{C}}$, the pseudomode approach reproduces the same QME as the Born--Markov treatment in Ref.~\cite{yamaguchi2021theory}, even in the presence of large detuning.
Equivalently, the auxiliary system may be modeled as consisting of two cavity modes, one of which (the background mode) exhibits an extremely large leakage rate and can therefore be adiabatically eliminated
\cite{garraway1997nonperturbative,bay1998superradiance, dalton2001theory, ben2026interference, lu2022unveiling}.

Second, we show that the constant term $J_0$ is necessary to make the QME of Lindblad form.
To see this, we write the QME in Eq.~\eqref{the_QME_via_Pseudomode_approach} as
\begin{align}
  \frac{d}{dt} \hat{\rho}(t)
   & = - i [ \hat{H}_{\mathrm{AC}} , \hat{\rho}(t) ]
  \\
   & + \sum_{m,n=1}^2 \Gamma_{m,n} \Big( \hat{X}_{m} \hat{\rho}(t) \hat{X}_{n}^{\dagger} - \frac{1}{2} \big\{ \hat{X}_{n}^{\dagger} \hat{X}_{m} , \hat{\rho}(t) \big\} \Big),
  \notag
\end{align}
where $\hat{X}_{1} = \hat{\sigma}$ and $\hat{X}_{2} = \hat{a}$
and the Kossakowski matrix $\Gamma$ is given by
\begin{align}
  \Gamma =
  \begin{pmatrix}
    2 \pi J_0 & 2 \nu^*       \\
    2 \nu     & - 2 \Im z_{1}
  \end{pmatrix}.
\end{align}
It is known that the QME is of Lindblad form if and only if $\Gamma$ is positive semidefinite \cite{gorini1976completely,lindblad1976generators,breuer2002theory}.
Therefore, the following condition is necessary:
\begin{align}
  \label{condition_for_Lindblad_form}
  (- \Im z_{1}) \pi J_0 \ge |\nu|^{2}.
\end{align}
Within the correspondence in Eqs.~\eqref{correspondence_alpha}--\eqref{correspondence_g1} with $\gamma_{\mathrm{F}} = \sqrt{\eta \gamma \kappa} e^{i(\theta_{\mathrm{A}} - \theta_{\mathrm{C}})}$ and $0 \le \eta \le 1$, this condition is satisfied and therefore Yamaguchi's QME~\eqref{goal_of_the_QME} is of Lindblad form.
The condition~\eqref{condition_for_Lindblad_form} shows the importance of the constant term $J_0$ for the QME to be of Lindblad form.
In broader perspective, this gives a method for modifying a non-Lindblad QME to be of Lindblad form:
if a QME obtained in the pseudomode approach is of non-Lindblad form, adding a constant term in the spectral function can convert it into a Lindblad form.
This method is contrastive to the standard modification of QME by means of a rotation of pseudomodes \cite{garraway1997nonperturbative, pleasance2020generalized}.
A rotation of pseudomodes may transform a QME into Lindblad form, while it generally gives rise to additional interactions between the pseudomodes.


\section{Derivation of the spectral function via Fano diagonalization}\label{sec:Fano_diagonalization}

In this section, we derive the form of the spectral function $J(\omega)$ directly from the atom-cavity-reservoir setup in Ref.~\cite{yamaguchi2021theory}, 
where the cavity mode is treated explicitly as part of the system of interest [Fig.~\ref{fig:setup}(b)]. 
This approach is complementary to the pseudomode approach adopted in Secs.~\ref{sec:setup} and \ref{sec:pseudomode_approach}.
Whereas the pseudomode approach identifies the spectral function required to reproduce the QME \eqref{goal_of_the_QME}, 
the derivation presented here starts from the atom-cavity-reservoir model and incorporates the cavity degree of freedom into the reservoir through Fano diagonalization.

We first derive the spectral function for the case of perfect interference ($\eta = 1$), for which the atom and cavity couple to a single common reservoir and the cavity and the reservoir are Fano-diagonalized, yielding Eq.~\eqref{spectral_function_eta_1}. 
We then generalize the derivation to the case of imperfect interference ($0 < \eta < 1$), where the atom and cavity share only part of the reservoir degrees of freedom and the description requires an additional independent reservoir coupled exclusively to the atom, leading to Eq.~\eqref{spectral_function_e_q}.

\subsection{Case of perfect interference ($\eta = 1$)}\label{subsec:perfect_case}

We first derive $J(\omega)$ for the perfect interference case ($\eta = 1$).
We assume a single reservoir that is a one-dimensional bosonic system with modes labeled by the mode frequency $\omega$.
The reservoir's annihilation and creation operators, denoted by $\hat{A}_{\omega}$ and $\hat{A}_{\omega}^{\dagger}$, satisfy $[\hat{A}_{\omega} , \hat{A}_{\omega'}^{\dagger}] = \delta(\omega - \omega')$.

The total Hamiltonian for the setup in Fig.~\ref{fig:setup}(b) is given by $\hat{H}' = \hat{H}_0 + \hat{H}_{\mathrm{R}1} + \hat{H}_{\mathrm{I}1}$.
$\hat{H}_0$ is the Hamiltonian of the system of the atom and the cavity and is defined by Eq.~\eqref{goal_of_the_QME_Hamiltonian_part}.
The reservoir's free Hamiltonian $\hat{H}_{\mathrm{R}1}$ and the system--reservoir interaction Hamiltonian $\hat{H}_{\mathrm{I}1}$ are defined by
\begin{align}
  \hat{H}_{\mathrm{R}1} & = \int d\omega \, \omega \hat{A}_{\omega}^{\dagger} \hat{A}_{\omega},
  \label{H_R1}
  \\
  \hat{H}_{\mathrm{I}1} & = \int d\omega \, \bigl( \xi_{\omega} \hat{\sigma}^{\dagger} \hat{A}_{\omega} + \xi_{\omega}^{*} \hat{\sigma} \hat{A}_{\omega}^{\dagger}
  + \zeta_{\omega} \hat{a}^{\dagger} \hat{A}_{\omega} + \zeta_{\omega}^{*} \hat{a} \hat{A}_{\omega}^{\dagger} \bigr).
  \label{H_I1}
\end{align}
Similarly to the treatment in Ref.~\cite{yamaguchi2021theory}, we assume that the atom--reservoir coupling strength $\xi_{\omega}$ and the cavity--reservoir coupling strength $\zeta_{\omega}$ are independent of $\omega$, so that
\begin{align}
  \xi_{\omega} = \sqrt{\frac{\gamma}{2\pi}}  e^{i\theta_{\mathrm{A}}},
  \quad
  \zeta_{\omega} = \sqrt{\frac{\kappa}{2\pi}} e^{i\theta_{\mathrm{C}}}.
  \label{coupling_for_perfect}
\end{align}

We employ the Fano diagonalization method \cite{fano1961effects,dalton2001theory} to rewrite the total Hamiltonian $\hat{H}'$ in a form where the atom is coupled to the Fano-diagonalized modes of the cavity--reservoir system.
This procedure enables us to derive the spectral function for the atom in this setup.

To this end, we first decompose the total Hamiltonian as $\hat{H}' = \hat{H}_{\mathrm{A}} + \hat{H}_{\mathrm{F}} + \hat{H}_{\mathrm{AF}}$,
where the atom's free Hamiltonian $\hat{H}_{\mathrm{A}} = \omega_{\mathrm{A}} \hat{\sigma}^{\dagger} \hat{\sigma}$ is the same as that in Eq.~\eqref{H_system_setup}, and the other two Hamiltonians are given by
\begin{align}
  \label{H_Fano_setup}
  \hat{H}_{\mathrm{F}}
   & = \omega_{\mathrm{C}} \hat{a}^{\dagger} \hat{a} + \int d\omega \, \omega \hat{A}_{\omega}^{\dagger} \hat{A}_{\omega}
  \notag
  \\
   & \quad + \sqrt{\frac{\kappa}{2 \pi}} \int d\omega \, \bigl( e^{i \theta_{\mathrm{C}}} \hat{a}^{\dagger} \hat{A}_{\omega} + e^{-i \theta_{\mathrm{C}}} \hat{a} \hat{A}_{\omega}^{\dagger} \bigr),
  \\
  \label{H_interaction_atom_and_Fano_setup}
  \hat{H}_{\mathrm{AF}}
   & = g \hat{\sigma}^{\dagger} \hat{a} + g^* \hat{\sigma} \hat{a}^{\dagger}
  \notag
  \\
   & \quad + \sqrt{\frac{\gamma}{2\pi}} \int d\omega \, \bigl( e^{i \theta_{\mathrm{A}}} \hat{\sigma}^{\dagger} \hat{A}_{\omega} + e^{-i \theta_{\mathrm{A}}} \hat{\sigma} \hat{A}_{\omega}^{\dagger} \bigr).
\end{align}

We then note that $\hat{H}_{\mathrm{F}}$ is quadratic in the bosonic operators $\hat{a}$ and $\hat{A}_{\omega}$,
Therefore, we can be diagonalize it by introducing an operator $\hat{B}_{\omega}$ defined by
\begin{align}
  \label{def_Fano_diagonalization_mode_operator}
  \hat{B}_{\omega} = \alpha(\omega) \hat{a} + \int d\omega' \, \beta(\omega,\omega') \hat{A}_{\omega'},
\end{align}
such that $\hat{B}_{\omega}$ corresponds to an eigenmode of $\hat{H}_{\mathrm{F}}$:
\begin{align}
  [ \hat{B}_{\omega} , \hat{H}_{\mathrm{F}} ] = \omega \hat{B}_{\omega}.
\end{align}
We also impose the orthonormal condition:
\begin{align}
  [ \hat{B}_{\omega} , \hat{B}_{\omega'}^{\dagger} ] = \delta(\omega-\omega').
\end{align}
From these equations, we obtain
\begin{align}
  \alpha(\omega)
   & = \sqrt{\frac{\kappa}{2 \pi}} \frac{e^{-i \theta_{\mathrm{C}} + i \psi}}{\omega - \omega_{\mathrm{C}} - i \kappa / 2},
  \\
  \beta(\omega,\omega')
   & = \frac{e^{i \psi}}{\omega - \omega_{\mathrm{C}} - i \kappa / 2 } \Bigl[ \frac{\kappa}{2 \pi} \frac{\mathcal{P}}{ \omega - \omega' } + ( \omega - \omega_{\mathrm{C}} ) \delta(\omega - \omega') \Bigr].
\end{align}
Here, $\psi$ is an arbitrary phase.
We have used the result of the principal-value integral $\int d\omega' \, \mathcal{P} / (\omega - \omega') = 0$.

According to the definition of $\hat{B}_{\omega}$ in Eq.~\eqref{def_Fano_diagonalization_mode_operator},
the operators $\hat{a}$ and $\hat{A}_{\omega}$ can be written as
\begin{align}
  \label{a_c_and_b_k_from_Fano_diagonalization_mode}
  \hat{a}
   & = \int d\omega \, \alpha^{*}(\omega) \hat{B}_{\omega},
  \\
  \hat{A}_{\omega}
   & = \int d\omega' \, \beta^{*}(\omega' , \omega) \hat{B}_{\omega'}.
\end{align}
By substituting these expressions into Eqs.~\eqref{H_Fano_setup} and \eqref{H_interaction_atom_and_Fano_setup}, we can rewrite the total Hamiltonian $\hat{H}'$ as
\begin{align}
  \hat{H}'
   & = \omega_{\mathrm{A}} \hat{\sigma}^{\dagger} \hat{\sigma}
  + \int d\omega \, \omega \hat{B}_{\omega}^{\dagger} \hat{B}_{\omega}
  \notag
  \\
   & \quad + \int d\omega \, \bigl[ \Lambda(\omega) \hat{\sigma}^{\dagger} \hat{B}_{\omega} + \Lambda^{*}(\omega) \hat{\sigma} \hat{B}_{\omega}^{\dagger} \bigr],
  \label{Fano_diagonalized_H}
\end{align}
where the second and third terms in the right-hand side result from $\hat{H}_{\mathrm{F}}$ and  $\hat{H}_{\mathrm{AF}}$, respectively.
Here, $\Lambda(\omega)$ represents the coupling strength between the atom and the Fano-diagonalized mode $\hat{B}_{\omega}$ and is given by
\begin{align}
  \label{coupling_strength_Fano}
  \Lambda(\omega)
   & = g \alpha^{*}(\omega) + \sqrt{\frac{\gamma}{2 \pi}} e^{i \theta_{\mathrm{A}}} \int d\omega' \, \beta^{*}(\omega', \omega),
  \\
   & = \frac{e^{-i \psi}}{\omega - \omega_{\mathrm{C}} + i \kappa / 2 }
  \biggl[ g \sqrt{ \frac{\kappa}{2 \pi} } e^{i \theta_{\mathrm{C}}} + (\omega - \omega_{\mathrm{C}}) \sqrt{ \frac{\gamma}{2 \pi} } e^{i \theta_{\mathrm{A}}} \biggr].
  \notag
\end{align}
The total Hamiltonian in Eq.~\eqref{Fano_diagonalized_H} describes the setup in Fig.~\ref{fig:setup}(a), in which the atom interacts with the structured reservoir obtained through Fano diagonalization.

From Eq.~\eqref{coupling_strength_Fano}, we obtain the spectral function $J(\omega) = |\Lambda(\omega)|^{2}$ for the atom--reservoir coupling and easily find that it coincides with Eq.~\eqref{spectral_function_eta_1}, the spectral function for the case of $\eta = 1$ obtained within the pseudomode approach.

We note that the direct atom--environment coupling, i.e., $\xi_{\omega} \neq 0$ ($\gamma \neq 0$) is essential for generating the Fano interference between the radiation directly from the atom and that mediated by the cavity.
In contrast to the present study, the standard Fano diagonalization method \cite{dalton2001theory} assumes the absence of the direct coupling, i.e., $\xi_{\omega} = 0$ ($\gamma = 0$).
In such a case of the atom--cavity system, Eq.~\eqref{coupling_strength_Fano} shows that $\Lambda(\omega)$ does not involve the term depending on $\beta(\omega, \omega')$, and consequently no Fano effect arises.
If one wishes to discuss Fano effect while retaining the absence of direct coupling, at least two cavity modes are required \cite{dalton2001theory}.

\subsection{Case of imperfect interference ($\eta < 1$)}\label{subsec:imperfect_case}

We next derive $J(\omega)$ for the imperfect interference case ($0 < \eta < 1$).
In the derivation for the perfect case, the spectral function is given by the modulus of the (generally complex) coupling $\Lambda(\omega)$ between the atom and the Fano-diagonalized reservoir.
In contrast, for the imperfect case, the spectral function given in Eq.~\eqref{spectral_function_e_q} is difficult to be expressed in such a simple form.
To overcome this difficulty, we introduce an additional reservoir into the setup used for the perfect case.
By taking into account the contribution of the coupling between the atom and this additional reservoir to the spectral function,
we can derive Eq.~\eqref{spectral_function_e_q} as shown below.

We assume that the additional reservoir is also a one-dimensional bosonic system. 
Its annihilation and creation operators are denoted by $\hat{C}_{\omega}$ and $\hat{C}_{\omega}^{\dagger}$, respectively, and are assumed to be independent of the cavity mode and the mode $\hat{A}_{\omega}$.
Accordingly, they satisfy $[\hat{C}_{\omega}, \hat{C}_{\omega'}^{\dagger}] = \delta(\omega - \omega')$ 
and $[\hat{a}, \hat{C}_{\omega'}] = [\hat{a}, \hat{C}_{\omega'}^{\dagger}] = [\hat{A}_{\omega}, \hat{C}_{\omega'}] = [\hat{A}_{\omega}, \hat{C}_{\omega'}^{\dagger}] = 0$.

The total Hamiltonian in the presence of the additional reservoir is given by 
$\hat{H}'' = \hat{H}' + \hat{H}_{\mathrm{R}2} + \hat{H}_{\mathrm{I}2}$,
where $\hat{H}' = \hat{H}_0 + \hat{H}_{\mathrm{R}1} + \hat{H}_{\mathrm{I}1}$ is the same as in the perfect interference case [Eqs.~\eqref{goal_of_the_QME_Hamiltonian_part}, \eqref{H_R1} and \eqref{H_I1}]. 
We assume that the Hamiltonians associated with the additional reservoir (the $\hat{C}_{\omega}$-mode reservoir) are given by
\begin{align}
  \hat{H}_{\mathrm{R}2} & = \int d\omega \, \omega \hat{C}_{\omega}^{\dagger} \hat{C}_{\omega},
  \\
  \hat{H}_{\mathrm{I}2} & = \int d\omega \, \bigl( \chi_{\omega} \hat{\sigma}^{\dagger} \hat{C}_{\omega} + \chi_{\omega}^* \hat{\sigma} \hat{C}_{\omega}^{\dagger} \bigr),
\end{align}
and that the coupling constants are given by
\begin{align}
  \xi_{\omega}  & = \sqrt{\eta} \sqrt{\frac{\gamma}{2\pi}}  e^{i\theta_{\mathrm{A}}},
  \quad
  \zeta_{\omega} = \sqrt{\frac{\kappa}{2\pi}} e^{i\theta_{\mathrm{C}}},
  \\
  \chi_{\omega} & = \sqrt{1 - \eta} \sqrt{\frac{\gamma}{2\pi}}  e^{i\varphi_{\mathrm{A}}},
  \label{chi_omega}
\end{align}
Compared with the couplings in the perfect interference case [Eq.~\eqref{coupling_for_perfect}], 
$\xi_{\omega}$ is reduced by a factor of $\sqrt{\eta}$. 
The Hamiltonian $\hat{H}''$ indicates that the atom is coupled to both the $\hat{A}_{\omega}$-mode reservoir and the $\hat{C}_{\omega}$-mode reservoir, 
whereas the cavity is coupled only to the $\hat{A}_{\omega}$-mode reservoir. 
The physical interpretation of this model is that a fraction $\eta$ of the atomic radiation is emitted into the common reservoir (the $\hat{A}_{\omega}$-mode reservoir), thereby participating in the Fano interference, while the remaining fraction $1-\eta$ is emitted into the independent reservoir (the $\hat{C}_{\omega}$-mode reservoir) and therefore does not contribute to the Fano effect.

Applying Fano diagonalization to $\hat{H}'$ in the same manner as in the perfect interference case,
we can rewrite $\hat{H}''$ as
\begin{align}
  \hat{H}''
   & = \omega_{\mathrm{A}} \hat{\sigma}^{\dagger} \hat{\sigma}
  + \hat{H}_{\mathrm{R}2} + \hat{H}_{\mathrm{I}2}
  + \int d\omega \, \omega \hat{B}_{\omega}^{\dagger} \hat{B}_{\omega}
  \notag
  \\
   & \quad + \int d\omega \, \bigl[ \Lambda(\omega; \eta) \hat{\sigma}^{\dagger} \hat{B}_{\omega} + \Lambda^{*}(\omega; \eta) \hat{\sigma} \hat{B}_{\omega}^{\dagger} \bigr],
\end{align}
where the Fano-diagonalized mode $\hat{B}_\omega$ is again given by Eq.~\eqref{def_Fano_diagonalization_mode_operator}
and the coupling between the atom and $\hat{B}_\omega$ becomes
\begin{align}
  & \Lambda(\omega; \eta)
  \notag\\
  & = \frac{e^{-i \psi}}{\omega - \omega_{\mathrm{C}} + i \kappa / 2 }
  \biggl[ g \sqrt{ \frac{\kappa}{2 \pi} } e^{i \theta_{\mathrm{C}}} + (\omega - \omega_{\mathrm{C}}) \sqrt{ \frac{\eta \gamma}{2 \pi} } e^{i \theta_{\mathrm{A}}} \biggr],
  \label{coupling_strength_Fano_eta}
\end{align}
where the second term is reduced by a factor of $\sqrt{\eta}$ compared with Eq.~\eqref{coupling_strength_Fano}.

Since the $\hat{B}_{\omega}$-mode reservoir and the $\hat{C}_{\omega}$-mode reservoir are independent,
the spectral function for the coupling between the atom and the reservoirs is given by
$J(\omega) = |\Lambda(\omega; \eta)|^2 + |\chi_\omega|^2$.
From Eqs.~\eqref{chi_omega} and \eqref{coupling_strength_Fano_eta}, we easily find that this spectral function coincides with Eq.~\eqref{spectral_function_e_q}, the spectral function for the case of $0 < \eta < 1$ (as well as $\eta = 0, 1$) obtained within the pseudomode approach.


\section{Summary}\label{sec:summary}

In this paper, we have shown that the pseudomode approach can be used to derive a QME for a cavity QED system exhibiting a Fano effect, in which radiation emitted directly by the atom interferes with that mediated by the cavity.
We have determined the spectral function of the coupling between the atom and the structured reservoir that leads to this QME, and demonstrated that a constant term in the spectral function plays a crucial role.
Rather than treating this term as a mere background contribution, we have related it directly to the rate associated with the Fano effect.
As a result, the resulting QME takes a Lindblad form.

The pseudomode framework extended in this work provides a complementary perspective to other approaches based on the input-output formalism and Markovian embedding methods \cite{tamascelli2018nonperturbative, lentrodt2020ab} including the quantum Langevin equation \cite{vcernotik2019cavity}, the few-mode mapping \cite{ben2026interference}, and the quasinormal mode approach \cite{franke2020quantized, meschede2026quantum}. Dissipators in these approaches are assumed to act solely on auxiliary modes. In contrast, in the present approach, dissipators act directly on the system (the atom) as well as on the auxiliary mode (the pseudomode). Related forms of dissipators have appeared in studies of cascaded quantum systems \cite{gardiner2004quantum, propp2022describe} and in efficient simulations based on the pseudomode approach \cite{PhysRevB.110.195148,huang2026coupled}.

We have further shown that the corresponding spectral function can be also obtained through Fano diagonalization in the atom--cavity--reservoir setup. 
For perfect interference, a single reservoir coupled to both the atom and the cavity is sufficient. 
For imperfect interference, however, an additional reservoir coupled only to the atom is required, reflecting the fact that part of the atomic radiation does not contribute to the Fano interference.

In particular, $\eta$ can be understood as quantifying the overlap between the radiation patterns of the atom and the cavity. Such an interpretation may also offer a useful perspective in the context of experimental implementations. \cite{lyasota2022mode}

The present work has clarified the non-Markovian origin of the Fano effect encoded in the spectral function, within the pseudomode approach and Fano diagonalization.
This provides a foundation for investigating how additional non-Markovian features influence Fano interference in open quantum systems.
In the present work, the analysis is performed in the single-excitation sector and assumes that the atom--environment and atom--cavity interactions are well described within the rotating-wave approximation.
Recent developments in Markovian embedding approaches beyond the single-excitation regime
\cite{lambert2019modelling,pleasance2020generalized,pleasance2021pseudomode,menczel2024nonhermitian,PhysRevB.110.195148,huang2026coupled}
suggest that the connection between Fano interference and structured non-Markovian environments identified here may persist more generally.
Nevertheless, a rigorous extension remains an important future problem.
Furthermore, in the ultrastrong-coupling regime, counter-rotating terms and high-frequency spectral tails may significantly affect the effective dynamics. Extending the present framework to such regimes is also left for future work.


\begin{acknowledgments}
  We thank Makoto Yamaguchi for valuable discussions.
  This work was supported by JST CREST Grant No. JPMJCR25I3, JST SPRING Grant No. JPMJSP2167, and JSPS KAKENHI Grants No. JP23K23231 and No. JP22H05032.
\end{acknowledgments}


\appendix


\section{Derivation of the effective Hamiltonian}\label{appendix:derivation_of_H_eff}

In this Appendix, we derive the effective Hamiltonian in Eq.~\eqref{def_effective_Hamiltonian_pseudomode} from the dynamical equations~\eqref{eq_derivative_c_1_t_in_sch_pic_pseudo} and \eqref{eq_derivative_b_1_t_in_sch_pic_pseudo} for the amplitudes $c_1(t)$ and $b_1(t)$.

The state vector of the enlarged system is given by Eq.~\eqref{def_state_vector_of_total_system_in_Sch_pic_pseudomode}.
Since $c_0$ is time independent, taking the time derivative yields
\begin{align}
  \frac{d}{dt}\ket*{\tilde{\Psi}(t)}
  = \frac{d c_1(t)}{dt}\ket*{\tilde{\Psi}_{1}} + \frac{d b_1(t)}{dt}\ket*{\tilde{\Phi}_{1}}.
\end{align}
Substituting Eqs.~\eqref{eq_derivative_c_1_t_in_sch_pic_pseudo} and \eqref{eq_derivative_b_1_t_in_sch_pic_pseudo}, we obtain
\begin{align}
  \frac{d}{dt}\ket*{\tilde{\Psi}(t)}
  =
   & -i \Bigl[ \bigl( \omega_{\mathrm A}-i\pi J_0 \bigr) c_1(t) + \tilde{g}_- b_1(t) \Bigr] \ket*{\tilde{\Psi}_{1}}
  \notag
  \\
   & -i \Bigl[ \tilde{g}_+^{*} c_1(t) + z_1 b_1(t) \Bigr] \ket*{\tilde{\Phi}_{1}}.
\end{align}

On the other hand, if Eq.~\eqref{sch_eq_at_total_system_in_sch_pic_pseudomode} holds, the action of the effective Hamiltonian on the state vector must satisfy
\begin{align}
  \hat{H}_{\mathrm{eff}} \ket*{\tilde{\Psi}(t)}
   & = c_0 \hat{H}_{\mathrm{eff}} \ket*{\tilde{\Psi}_{0}}
  + c_1(t) \hat{H}_{\mathrm{eff}} \ket*{\tilde{\Psi}_{1}}
  + b_1(t) \hat{H}_{\mathrm{eff}} \ket*{\tilde{\Phi}_{1}}
  \notag
  \\
   & = \Bigl[ \bigl( \omega_{\mathrm A} -i\pi J_0 \bigr) c_1(t) + \tilde{g}_- b_1(t) \Bigr] \ket*{\tilde{\Psi}_{1}}
  \notag
  \\
   & \quad + \Bigl[ \tilde{g}_+^{*} c_1(t) + z_1 b_1(t) \Bigr] \ket*{\tilde{\Phi}_{1}}.
  \label{action_H_eff}
\end{align}

In the subspace spanned by $\bigl\{ \ket*{\tilde{\Psi}_{0}}, \ket*{\tilde{\Psi}_{1}}, \ket*{\tilde{\Phi}_{1}} \bigr\}$,
corresponding to the restriction to at most one excitation, the operators satisfy
\begin{align}
  \hat{\sigma}^{\dagger} \hat{\sigma}
   & = \ketbra*{\tilde{\Psi}_{1}}{\tilde{\Psi}_{1}},
  \quad
  \hat{a}^{\dagger} \hat{a}
  = \ketbra*{\tilde{\Phi}_{1}}{\tilde{\Phi}_{1}},
  \\
  \hat{\sigma}^{\dagger} \hat{a}
   & = \ketbra*{\tilde{\Psi}_{1}}{\tilde{\Phi}_{1}},
  \quad
  \hat{\sigma} \hat{a}^{\dagger}
  = \ketbra*{\tilde{\Phi}_{1}}{\tilde{\Psi}_{1}}.
\end{align}
From Eq.~\eqref{action_H_eff} with these relations, we find that the effective Hamiltonian is determined as
\begin{align*}
  \hat{H}_{\mathrm{eff}}
  = (\omega_{\mathrm A}-i\pi J_0)\hat{\sigma}^{\dagger}\hat{\sigma}
  + z_1\hat{a}^{\dagger}\hat{a}
  + \tilde g_-\hat{\sigma}^{\dagger}\hat{a}
  + \tilde g_+^{*}\hat{\sigma}\hat{a}^{\dagger},
\end{align*}
which corresponds to the first line of Eq.~\eqref{def_effective_Hamiltonian_pseudomode}.


\section{Detailed derivation of the QME}\label{sec:appendix}

In this appendix, we present a detailed derivation of Eq.~\eqref{the_QME_via_Pseudomode_approach}.
To this end, we first analyze the probability $\Pi_{0}(t)$ that the atom is in the ground state.
In the original setup given in Eqs.~\eqref{H_system_setup}--\eqref{H_interaction_setup},
$\Pi_{0}(t)$ is given by
\begin{align}
  \Pi_{0}(t)
   & = \mathrm{Tr}_{\mathrm{A}+\mathrm{R}} \bigl[ (\ketbra{0}_{\mathrm{A}}) \bigl( \ketbra{\Psi(t)} \bigr) \bigr]
  \notag
  \\
  \label{def_Pi_0_t_1}
   & = |c_{0}|^{2} + \sum_{k} |c_{k}(t)|^{2}
  \\
  \label{def_Pi_0_t_2}
   & = 1 - |c_{1}(t)|^{2}
\end{align}
where $\mathrm{Tr}_{\mathrm{A}+\mathrm{R}}$ is the trace over the atom--reservoir system,
and $\ket{\Psi(t)}$ is the atom--reservoir state given in Eq.~\eqref{def_total_state_vector_Psi_t}.
In the third line, we have used the the normalization condition $\braket{\Psi(t)} = 1$.
On the other hand, in the setup of the atom--pseudomode system, $\Pi_{0}(t)$ is given by
\begin{align}
  \Pi_{0}(t)
   & = \mathrm{Tr}_{\mathrm{A}+\mathrm{C}} \bigl[ (\ketbra{0}_{\mathrm{A}}) \{ \hat{\rho}_{\mathrm{nj}}(t) + \hat{\rho}_{\mathrm{j}}(t) \} \bigr]
  \notag
  \\
  \label{result_Pi_0_t_in_pseudomode}
   & = |c_{0}|^{2} + |b_{1}(t)|^{2} + \Pi_{\mathrm{j}}(t),
  %
\end{align}
where $\mathrm{Tr}_{\mathrm{A}+\mathrm{C}}$ is the trace over the atom--pseudomode system
and $\hat{\rho}_{\mathrm{nj}}(t)$ given in Eq.~\eqref{def_rho_nj_t} is the non-quantum-jump contribution to the density matrix,
and $\hat{\rho}_{\mathrm{j}}(t)$ given in Eq.~\eqref{def_rho_j_t} is the quantum-jump contribution.
Comparing Eqs.~\eqref{def_Pi_0_t_1} and \eqref{result_Pi_0_t_in_pseudomode},
we find that the term of $\sum_{k} |c_{k}(t)|^{2}$ in the original setup is decomposed into two contributions, $|b_{1}(t)|^{2}$ and $\Pi_{\mathrm{j}}(t)$, in the atom--pseudomode system.
Therefore, $\Pi_{\mathrm{j}}(t)$ can be interpreted as the probability that both the atom and the auxiliary system (pseudomode) are in their vacuum states while the remaining reservoir system (except for the auxiliary system) is in the singly-excited states.
It can be also interpreted as a correction term that ensures that the probability $\Pi_{0}(t)$ is consistently expressed within the pseudomode approach in terms of $c_{0}$, $c_{1}(t)$, and $b_{1}(t)$.

We next calculate the time derivative of $\Pi_{\mathrm{j}}(t)$.
From Eqs.~\eqref{def_Pi_0_t_2} and \eqref{result_Pi_0_t_in_pseudomode},
$\Pi_{\mathrm{j}}(t)$ is expressed as
\begin{align}
  \label{Pi_j_t_in_pseudomode}
  \Pi_{\mathrm{j}}(t) = 1 - |c_{0}|^{2} - |c_{1}(t)|^{2} - |b_{1}(t)|^{2}.
\end{align}
Using Eq.~\eqref{eq_derivative_c_1_t_in_sch_pic_pseudo}, we calculate the derivative of $|c_{1}(t)|^{2}$ as
\begin{align}
  \frac{d}{dt} |c_{1}(t)|^{2}
   & = -2 \pi J_0 |c_{1}(t)|^{2}
  + i \tilde{g}_-^{*} c_{1}(t) b_{1}^{*}(t)
  - i \tilde{g}_- c_{1}^{*}(t) b_{1}(t).
  \notag
\end{align}
Using Eq.~\eqref{eq_derivative_b_1_t_in_sch_pic_pseudo}, we calculate the derivative of $|b_{1}(t)|^{2}$ as
\begin{align}
  \frac{d}{dt} |b_{1}(t)|^{2}
   & = 2 \Im z_{1} |b_{1}(t)|^{2}
  - i \tilde{g}_+^{*} c_{1}(t) b_{1}^*(t)
  + i \tilde{g}_+ c_{1}^*(t) b_{1}(t).
  \notag
\end{align}
We thus obtain
\begin{align}
  \frac{d}{dt} \Pi_{\mathrm{j}}(t)
   & = 2 \pi J_0 |c_{1}(t)|^{2} - 2 \Im z_{1} |b_{1}(t)|^{2}
  \notag
  \\
  \label{eq_derivative_Pi_j_t}
   & \quad + 2 \bigl[ \nu c_{1}^{*}(t) b_{1}(t) + \nu^* c_{1}(t) b_{1}^{*}(t) \bigr],
\end{align}
where we have used $\tilde{g}_\pm = \mu \pm i \nu$.
We retain $\tilde{g}_\pm$ to be complex-valued for deriving the QME,
in contrast to the standard approach that resorts to a rotation of pseudomodes \cite{garraway1997nonperturbative}.

Furthermore, by noting the following relations
\begin{align*}
  |c_{1}(t)|^{2}
   & = \bra{\tilde{\Psi}_{0}} \hat{\sigma} \hat{\rho}_{\mathrm{nj}}(t) \hat{\sigma}^{\dagger} \ket{\tilde{\Psi}_{0}},
  \\
  |b_{1}(t)|^{2}
   & = \bra{\tilde{\Psi}_{0}} \hat{a} \hat{\rho}_{\mathrm{nj}}(t) \hat{a}^{\dagger} \ket{\tilde{\Psi}_{0}},
  \\
  c_{1}^{*}(t) b_{1}(t)
   & = \bra{\tilde{\Psi}_{0}} \hat{a} \hat{\rho}_{\mathrm{nj}}(t) \hat{\sigma}^{\dagger} \ket{\tilde{\Psi}_{0}},
  \\
  c_{1}(t) b_{1}^{*}(t)
   & = \bra{\tilde{\Psi}_{0}} \hat{\sigma} \hat{\rho}_{\mathrm{nj}}(t) \hat{a}^{\dagger} \ket{\tilde{\Psi}_{0}},
\end{align*}
we have
\begin{align}
  \label{c_1_t_2}
  |c_{1}(t)|^{2} \ketbra{\tilde{\Psi}_{0}}
   & = \hat{\sigma} \hat{\rho}_{\mathrm{nj}}(t) \hat{\sigma}^{\dagger},
  \\
  \label{b_1_t_2}
  |b_{1}(t)|^{2} \ketbra{\tilde{\Psi}_{0}}
   & = \hat{a} \hat{\rho}_{\mathrm{nj}}(t) \hat{a}^{\dagger},
  \\
  \label{c_1_b_1_1}
  c_{1}^{*}(t) b_{1}(t) \ketbra{\tilde{\Psi}_{0}}
   & = \hat{a} \hat{\rho}_{\mathrm{nj}}(t) \hat{\sigma}^{\dagger},
  \\
  \label{c_1_b_1_2}
  c_{1}(t) b_{1}^{*}(t) \ketbra{\tilde{\Psi}_{0}}
   & = \hat{\sigma} \hat{\rho}_{\mathrm{nj}}(t) \hat{a}^{\dagger}.
\end{align}
Then, combining Eqs.~\eqref{def_rho_j_t} and \eqref{eq_derivative_Pi_j_t}--\eqref{c_1_b_1_2}, we obtain the time derivative of $\hat{\rho}_{\mathrm{j}}(t)$:
\begin{align}
  \frac{d}{dt} \hat{\rho}_{\mathrm{j}}(t)
   & = 2 \pi J_0 \hat{\sigma} \hat{\rho}_{\mathrm{nj}}(t) \hat{\sigma}^{\dagger}
  - 2 \Im z_{1}  \hat{a} \hat{\rho}_{\mathrm{nj}}(t) \hat{a}^{\dagger}
  \notag
  \\
   & \quad + 2 \bigl[ \nu \hat{a} \hat{\rho}_{\mathrm{nj}}(t) \hat{\sigma}^{\dagger}
  + \nu^* \hat{\sigma} \hat{\rho}_{\mathrm{nj}}(t) \hat{a}^{\dagger} \bigr].
  \label{eq_derivative_rho_j_t}
\end{align}

Finally, by adding Eq.~\eqref{eq_derivative_rho_j_t} to Eq.~\eqref{eq_derivative_rho_nj_t}
and noting the following relations:
\begin{align*}
   & \hat{\sigma} \hat{\rho}_{\mathrm{j}}(t) \hat{\sigma}^{\dagger}
  = \hat{a} \hat{\rho}_{\mathrm{j}}(t) \hat{a}^{\dagger}
  = \hat{a} \hat{\rho}_{\mathrm{j}}(t) \hat{\sigma}^{\dagger}
  = \hat{\sigma} \hat{\rho}_{\mathrm{j}}(t) \hat{a}^{\dagger}
  = 0,
  \\
   & \hat{H}_{\mathrm{eff}} \hat{\rho}_{\mathrm{j}}(t)
  - \hat{\rho}_{\mathrm{j}}(t) \hat{H}_{\mathrm{eff}}^{\dagger}
  = 0,
\end{align*}
we obtain
\begin{align*}
  \frac{d}{dt} \hat{\rho}(t)
   & = - i \bigl( \hat{H}_{\text{eff}} \hat{\rho}(t)
  - \hat{\rho}(t) \hat{H}_{\text{eff}}^{\dagger} \bigr)
  + 2 \pi J_0 \hat{\sigma} \hat{\rho}(t) \hat{\sigma}^{\dagger}
  \notag
  \\
   & \quad   - 2 \Im z_{1} \hat{a} \hat{\rho}(t) \hat{a}^{\dagger}
  + 2 \bigl[ \nu \hat{a} \hat{\rho}(t) \hat{\sigma}^{\dagger}
    + \nu^* \hat{\sigma} \hat{\rho}(t) \hat{a}^{\dagger} \bigr].
\end{align*}
By rearranging the terms in $\hat{H}_{\text{eff}}$, we arrive at the desired QME~\eqref{the_QME_via_Pseudomode_approach}.


\bibliography{reference-2.bib}

\end{document}